\documentclass[runningheads]{llncs}
\usepackage{graphicx}
\usepackage{subfigure}
\usepackage[pdftex]{hyperref}
\usepackage{multirow}
\usepackage{amsmath}
\begin{document}

\title{Automated Detection of Coronary Artery Stenosis in X-ray Angiography using Deep Neural Networks}
\titlerunning{Automated Stenosis Detection}
%
\author{Dinis L. Rodrigues\inst{1} \and
Miguel Nobre Menezes\inst{2} \and
Fausto J. Pinto\inst{2} \and \\
Arlindo L. Oliveira\inst{1}}

\authorrunning{D. L. Rodrigues et al.}
%
\institute{Instituto Superior Técnico, University of Lisbon, Portugal \and
CCUL, Lisbon School of Medicine, University of Lisbon, Portugal
}

\maketitle
\begin{abstract}
Coronary artery disease leading up to stenosis, the partial or total blocking of coronary arteries, is a severe condition that affects millions of patients each year. Automated identification and classification of stenosis severity from minimally invasive procedures would be of great clinical value, but existing methods do not match the accuracy of experienced cardiologists, due to the complexity of the task. Although a number of computational approaches for quantitative assessment of stenosis have been proposed to date, the performance of these methods is still far from the required levels for clinical applications. 
In this paper, we propose a two-step deep-learning framework to partially automate the detection of stenosis from X-ray coronary angiography images. In the two steps, we used two distinct convolutional neural network architectures, one to automatically identify and classify the angle of view, and another to determine the bounding boxes of the regions of interest in frames where stenosis is visible. Transfer learning and data augmentation techniques were used to boost the performance of the system in both tasks. We achieved a 0.97 accuracy on the task of classifying the Left/Right Coronary Artery (LCA/RCA) angle view and 0.68/0.73 recall on the determination of the regions of interest, for LCA and RCA, respectively. These results compare favorably with previous results obtained using related approaches, and open the way to a fully automated method for the identification of stenosis severity from X-ray angiographies.

\end{abstract}
\section{Introduction and Related Work}
\label{sec:sota}
Coronary artery disease (CAD), characterized by plaque buildup inside the coronary arteries, is the leading non-communicable disease in global mortality. This buildup leads to stenosis, the partial or total  blocking of blood flow in the coronary arteries leading to improper delivery of oxygen-rich blood to the heart, weakening the heart muscle, and possibly leading to heart failure. Current standard diagnosis methods rely on an expert physician to assess the case, off or on-site, using non-invasive or invasive procedures\cite{Bio:CADd,Coro:stat}. Although significant resources have been invested in prevention, proper available computer-assisted assessment and treatment procedures are still not available.

The availability of data and access to more computing power opened a path to a number of new convolutional neural network (CNN) designs and applications. The ResNet-50 is a popular architecture for image classification, introducing shortcut connections between layers, which allow the flow of relevant features into deeper layers \cite{imp:resnetpaper}. Based on the single-shot detector architecture for object detection, the RetinaNet \cite{imp:retinanet} combines the ResNet-50 with feature pyramid networks (FPN) \cite{imp:fpn}, assembling a richer backbone feature extractor with two additional sub-networks. These sub-networks are responsible for correctly classifying a bounding box and regressing the estimated coordinates.

Early developments in deep learning techniques applied in cardiology imaging began by focusing on the structural segmentation of the ventricles and stenosis centerline extraction from Magnetic Resonance Imaging (MRI) and computed-tomography (CT) scans with the objective of aiding expert physicians in visual assessment \cite{sota:mri,sota:mri3,sota:mri4,sota:mri2,mri:5}.

To deal with the lack of public Invasive Coronary Angiography (ICA) datasets  a custom CNN was trained on thousands of artificial 32 by 32 pixel patches \cite{sota:stenDet2018}  mimicking the presence of stenosis. The use of a sliding window with the patches dimensions on the original frame with the trained CNN led to increased detection performance but real test images were very few and were also scaled down.

To automate the process from start to finish the work by Au and co-authors \cite{AI:sten1} showcased a pipeline composed by three uniquely designed CNNs with the intention of detecting, segmenting and classifying stenosis severity through Quantitative Coronary Angiography (QCA) annotations in ICA reference images of the left coronary artery (LCA). Their study included 1024 study participants using only RCA viewing angles and reference frames. A detector variant of the single-shot detector YOLO \cite{AI:yolo} was developed with the objective of determining fixed dimensional regions of 192 by 192 pixels on which a stenosis was present. Within the proposed region another custom segmentation deep learning architecture was built, based on U-Net \cite{AI:unet}, to automatically segment every pixel where the stenosis was present. An additional small custom CNN with only five convolutional layers was built to classify the segmented frame.

Focusing only in the detection task of the stenosis Wu et al.  developed a novel single-shot architecture using the VGG16, a feature extractor from which feature maps from low and high level convolutional layers are extracted \cite{sota:stenDet2020}. Those feature maps are then passed into a classification and regression sub-network, to estimate bounding box coordinates and the respective confidence scores.

Cong et al. \cite{AI:sten4} proposed a three-stage end-to-end workflow for stenosis characterization. The process of viewing angle selection is initialized with transfer learning and fine-tuning of the InceptionV3 architecture \cite{imp:inceptionv3}. Features extracted from the last convolutional layer are used to train a bi-directional long short-term memory (LSTM) network, taking advantage of the temporal dimension to extract the exact frame of the sequence corresponding to the reference frame. With the extracted frame another InceptionV3 is fine-tuned for a classification task of the stenosis assessment under QCA labels. The detection of the stenosis is then performed as a weakly supervised method by employing class activation maps using Grad-CAM \cite{sota:gradCam} to identify the most important regions based on the weights contribution for the respective frame classification result. These detections are then evaluated against expert physician manual annotations of 35 by 35 pixel bounding boxes. We compare our results directly with this work, which is the only one that addresses a directly comparable challenge.

\section{Medical Data}\label{sec:meddata}

The data was curated by one of the authors, Miguel Menezes, who is an expert cardiologist, and was obtained from patients treated in the largest tertiary university hospital in Portugal. The data is composed of 9378 clinically obtained invasive coronary angiography sequences of 438 patients, in DICOM format, obtained between 2015 and 2019. The data was properly anonymized to preserve participant privacy and was authorized by the ethics committee. All participants were over the age of eighteen. 

For each subject, the value of the Instantaneous Flow Ratio (iFR) and the the stenoses location was included at the patient level. Annotations for the optimal sequences, i.e. those where stenosis is best seen under the radio-opaque contrast, were included in a non-destructive way using \textit{Osirix} \cite{meddata:osirix}, an image processing software.

For each coronary artery containing a stenosis: (1) a unique frame was annotated showcasing the artery; (2) a unique frame was annotated showcasing how the wire of the iFR measuring procedure was placed; (3) a unique frame was annotated showcasing all the stenosis of the corresponding artery. The unique frame annotated in (3) corresponds to the best contrast viewing frame and is considered the reference frame of the sequence. A total of 1593 sequences accounting for 438 patients were annotated using this procedure.

Only sequences with frame dimensions of 512 by 512 pixels in gray scale were used. To alleviate the labour of having to annotate all the bounding boxes in every single frame manually, the reference frame annotations were propagated to the entire sequence through the implementation of the object tracking algorithm \textit{Discriminative Correlation Filter Tracker with
Channel and Spatial Reliability} \cite{meddata:tracker}. Misplaced bounding boxes due to rapid shifts from frame to frame and/or occlusions were \textit{a posteriori} manually corrected to have a good fit to the stenosis.

\begin{table*}[!htp]
\centering
\caption{Processed dataset, with all optimal sequences, frame intervals, stenosis annotations, and iFR values count.}\label{meddata:stat}
\resizebox{\linewidth}{!}{\begin{tabular}{cccccccccc}
\hline
 &  &  & \multicolumn{4}{c}{Frames} &  &  &  \\ \cline{4-7}
Sequence Detail & Patients & Sequences & No Contrast & Introducing & Optimal & Vanishing & Stenosis Annotations & iFR below & iFR above \\ \hline
Total Sequences & 438 & 1593 & 0 & 0 & 0 & 0 & 4234 & 554 & 1005 \\
Discard & 72 & 115 & 0 & 0 & 0 & 0 & 338 & 40 & 73 \\
With Implants & 82 & 184 & 0 & 0 & 0 & 0 & 472 & 81 & 96 \\
Optimal & 392 & 1294 & 11582 & 1294 & 20819 & 39266 & 3424 & 433 & 836 \\
Optimal RCA & 91 & 235 & 2249 & 235 & 3983 & 6323 & 309 & 25 & 210 \\
Optimal LCA & 126 & 155 & 1323 & 155 & 2474 & 5077 & 225 & 70 & 85 \\
Optimal LCx/LAD & 111 & 118 & 1155 & 118 & 1912 & 3869 & 159 & 53 & 65 \\
Optimal LAD/LCx & 90 & 92 & 865 & 92 & 1590 & 2616 & 105 & 43 & 49 \\  
No Lesion RCA & 48 & 54 & 465	& 54 & 748 & 1450 & 0 & 0 & 0 \\
No Lesion LCA & 17 & 18 & 153 & 16 & 190 & 538 & 0 & 0 & 0
\end{tabular}}
\end{table*}

For each sequence, frame intervals were also labeled as: (a) No radio-opaque contrast; (b) The contrast is being introduced; (c) The contrast has been fully introduced (optimal frames); (d) The contrast is vanishing. Sequences were also grouped by their respective angles (obtained from the DICOM metadata) and then manually labeled to the most common viewing angle names. Sequences with metal implants, pacemakers and with the iFR medical suited wires were discarded. The 1593 annotated sequences from 498 patients were processed with the described steps, resulting in 1294 optimal sequences with 20819 optimal frames, where the bounding box was placed (see \autoref{meddata:stat}).
\section{Implementation}
\label{sec:imp}
\subsection{Angle View Selection}
The initial phases of detecting and assessing a stenosis require the identification of the coronary viewing angle. Since every sequence was previously labeled with the respective viewing angle, this task is addressed as a classification problem, since one frame can only belong to a specific viewing angle. The ResNet-50 \cite{imp:resnetpaper} was chosen as the architecture for this task.
\begin{figure}[!htp]
    \centering
    \includegraphics[width=\linewidth]{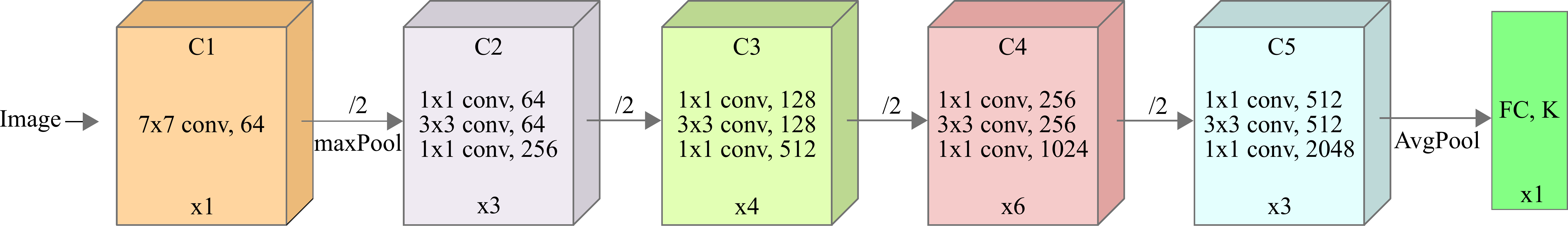}
    \caption{ResNet-50 simplified architecture. The bottom of each block denotes the repeated set of layers (Kernel size, operation, number of channels) in each convolution block ($C$). After each block ($C1$ to $C5$) the spatial dimension is reduced. The last layer is a fully connected layer with $K$ units. Shortcut connections occur every two layers but are omitted for readability.}
    \label{imp:res50}
\end{figure}
The ResNet-50 (see \autoref{imp:res50}) was trained and evaluated using 5-fold stratified cross-validation at sequence-level, for 30 epochs and with a batch size of 32. To improve convergence speed, stochastic gradient descent with the \textit{Adam} \cite{imp:adam} optimizer was performed. The initial learning rate was set at $\eta = 10^{-5}$, and reduced by a factor of $0.2$ on loss \textit{plateau}. To reduce early stages of overfitting and large gradient updates to the network, due to the weight initialization of the last fully connected layer, a two-stage training workflow was assembled as follows. For the first 15 epochs, the gradient updates on all layers of the network are frozen except for the $C_5$ block and fully connected block, stopping the gradient updates from becoming too large and preventing overfitting in early steps. For the next 15 epochs, the gradient updates of the entire model are restored allowing the model to converge in its entirety.

\subsection{Stenosis Detection}
The objective of this step is to detect and estimate the position of every visible stenosis in a given frame. Given the annotated bounding boxes for the stenosis in the optimal interval, it is possible to approximate this to an object detection/recognition problem where the stenosis is the object of interest.

\begin{figure}[!ht]
    \centering
    \includegraphics[width=\linewidth]{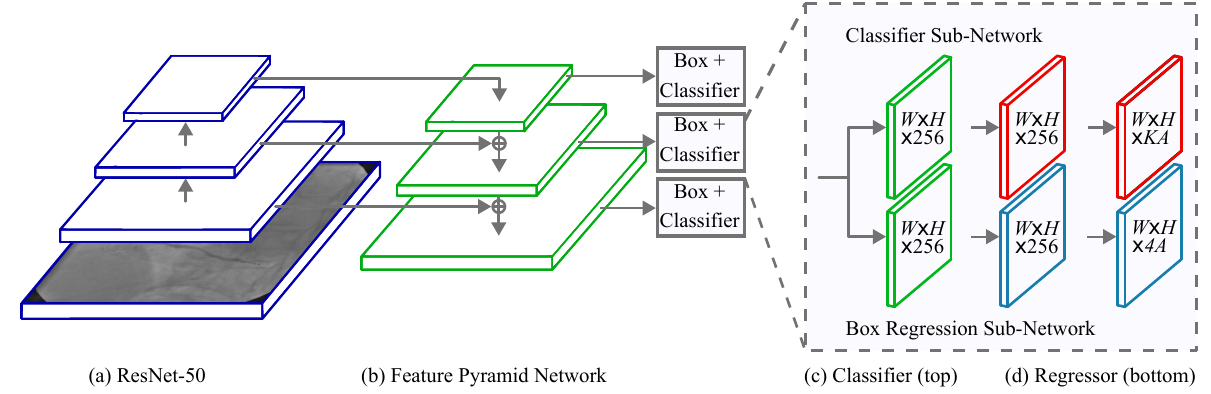}
    \caption{RetinaNet architecture with (a) ResNet-50 and (b) Feature Pyramid Network as feature extractor to (c) classify the lesion existence probability and (d) regress the bounding box coordinates.}
    \label{imp:retinaArch}
\end{figure}

We decided to use the RetinaNet \cite{imp:retinanet} architecture and methodology for the stenosis detection task. For bounding box classification and regression, translation invariant anchor boxes are generated in each pixel of the resulting feature maps. To improve bounding box coverage, aspect ratios $\{1$:$1$, $1$:$2$, $2$:$1$, $4$:$1\}$ and scales $\{ 2^0, 2^{1/2}, 2^1\}$ are established for each anchor box, resulting in $A=12$ anchors per feature map pixel.

For each 512 by 512 pixel image, approximately 45 thousand anchor boxes are generated. To deal with the amount of generated boxes and occurring class imbalance between background and stenosis assignment, the $\alpha$-balanced Focal Loss function \cite{imp:retinanet}
was used with $\alpha = 0.25$ and $\gamma = 2$. These values were obtained by testing the performance of the system in the stenosis detection task.

The regression sub-networks estimate the relative offset between the predicted anchor $\hat A$ and the matched ground-truth bounding box $G$. A parameterized regression target $T$ is calculated \cite{imp:boxPar} for each matched pair $(\hat A, G)$ as
\begin{equation}
\begin{matrix}
    t_x = \left(G_x - \hat A_x\right) / \hat A_w\\
    t_y = \left(G_y - \hat A_y\right) / \hat A_y\\
    t_w = \log \left(G_w / \hat A_w\right)\\
    t_h = \log \left(G_h / \hat A_h\right)
\end{matrix}
\end{equation}
where $(t_x, t_y)$ denotes the center scaling invariant translation, and $(t_w, t_h)$ a logarithmic space translation of the estimated width and height anchor $\hat A$. The network is trained to estimate these parameterized coordinates offset $T$ under the $\textrm{smooth}_{L_1}$ loss function \cite{imp:fastrcnn}, making the model more robust to outlier detections.


At inference, the non-maximum suppression algorithm is applied with a score threshold of 0.5 and an intersection-over-union (IoU) threshold defined at 0.5.

The model was trained and evaluated under 5-fold stratified cross-validation at patient-level, for 3500 steps with a batch size of 32, under stochastic gradient descent with an initial learning rate of $\eta = 8.10^{-4}$ and a momentum term $\gamma = 0.9$. Since high weight values increase the chances of overfitting, $L_2$ regularization with a weight factor $\lambda=4.10^{-4}$ was used, applying a penalty for the networks weight values.

Data augmentation techniques were also implemented during training, enhancing the quality and enlarging the size of the dataset. Augmentation was performed by generating additional modified versions in brightness and contrast from the original frames, in order to to reduce overfitting and improve generalization.

All tests were performed using Tensorflow 2.2.0 and Python 3.7. Computations were performed on a Debian 10 operating system, equipped with 64GB of RAM, an 8-core 2.4GHz Intel Xeon Silver 4214R CPU, and a 32GB NVIDIA Tesla V100S GPU.

\section{Results}\label{sec:results}
\subsection{Angle selection performance}
From the original 512 by 512-pixel dimensions, a version of the input was generated by down-scaling it to 224x224. The objective was to understand what was the resolution better adapted to the task at hand.
\begin{figure}[!htp]
    \centering
    \subfigure[512 RCA]{\includegraphics[width=0.24\linewidth]{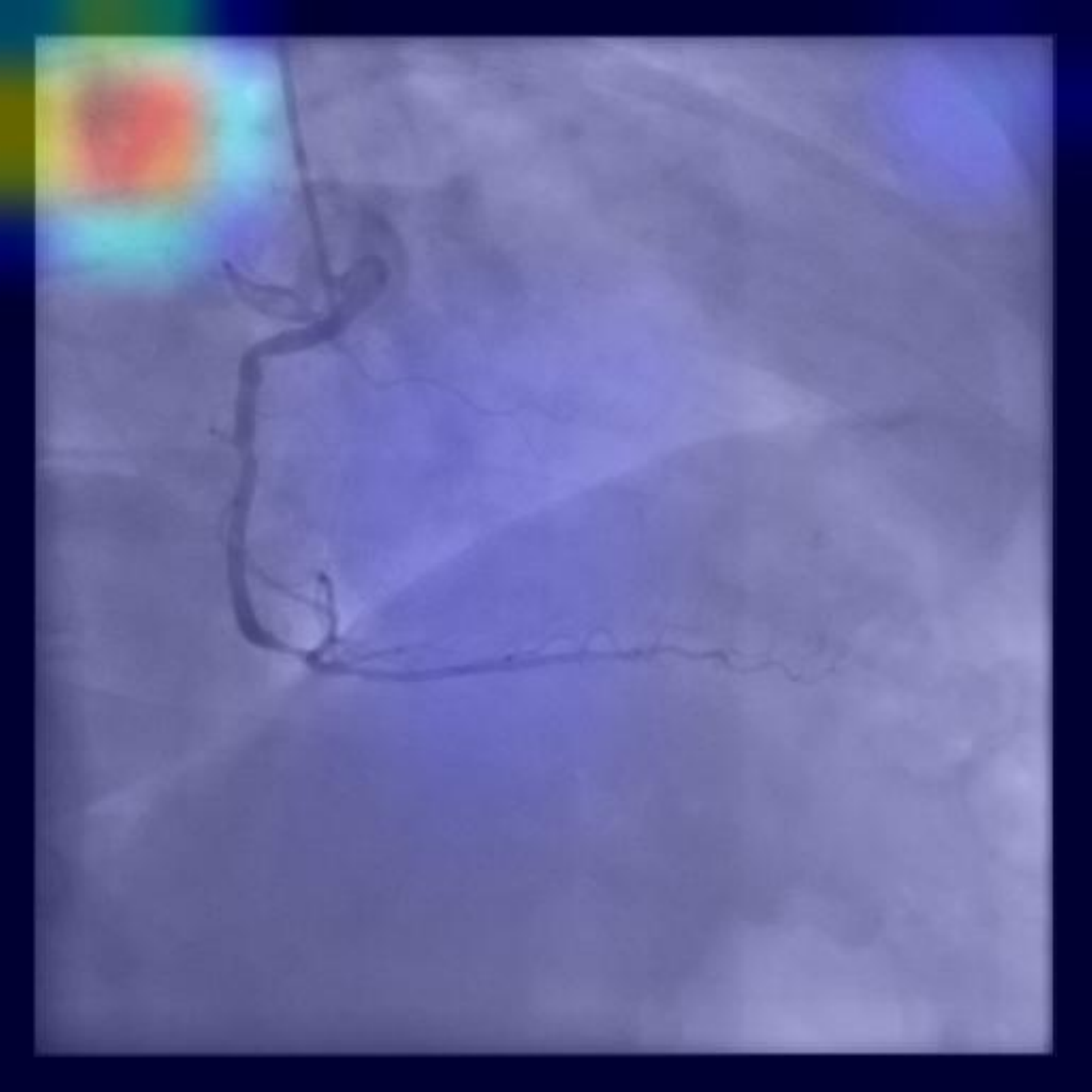}}
     \subfigure[224 RCA]{\includegraphics[width=0.24\linewidth]{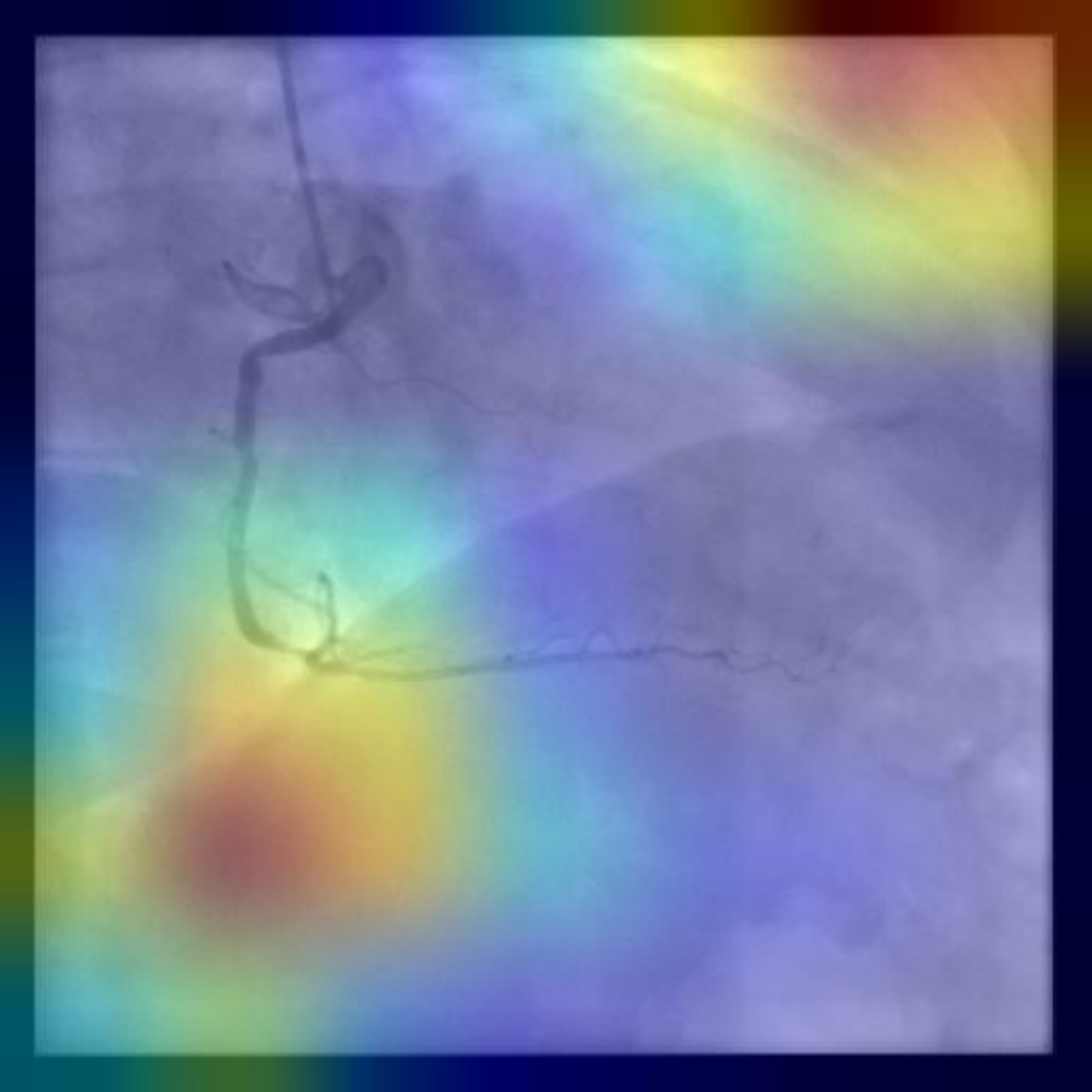}}
    \subfigure[512 LCA]{\includegraphics[width=0.24\linewidth]{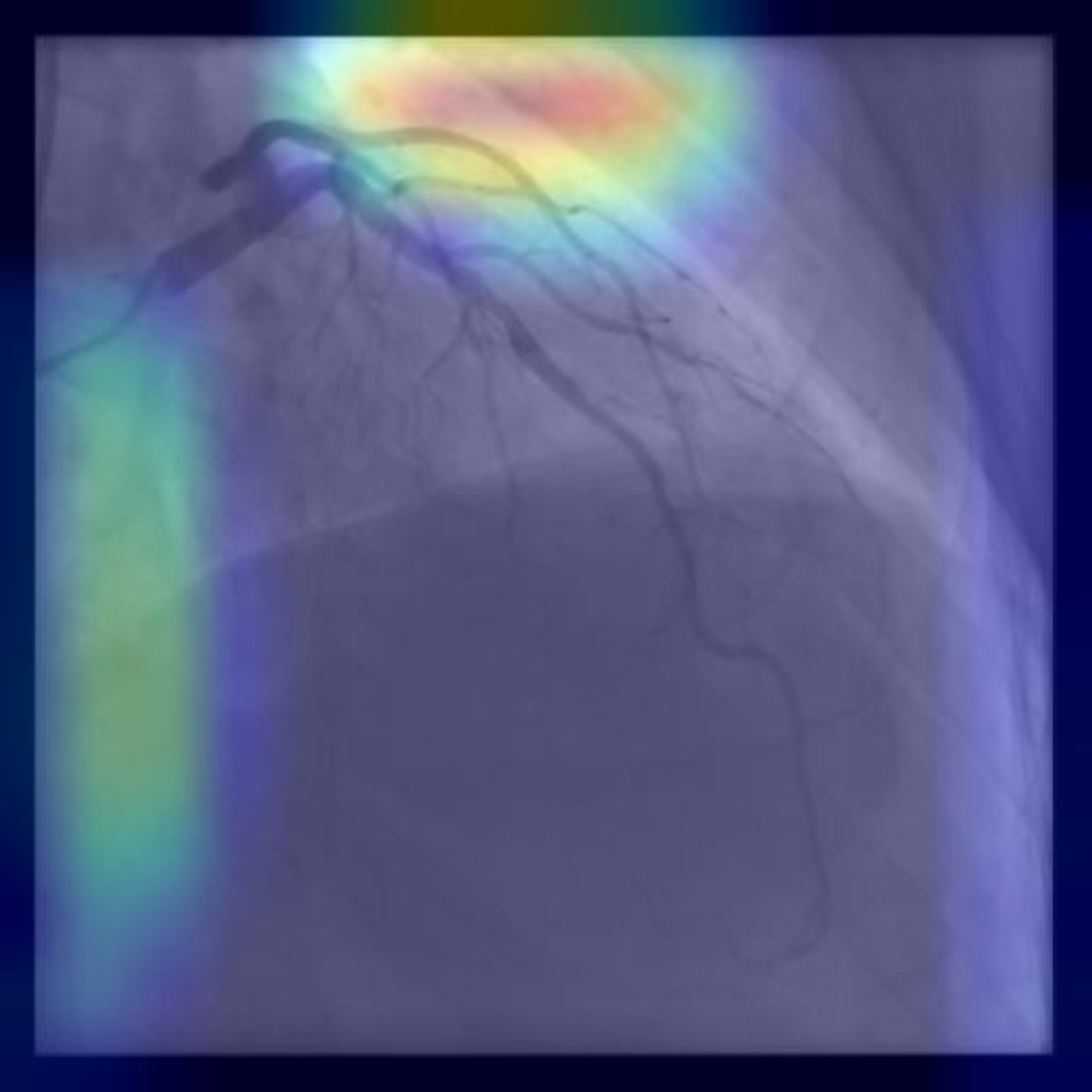}} 
    \subfigure[224 LCA]{\includegraphics[width=0.24\linewidth]{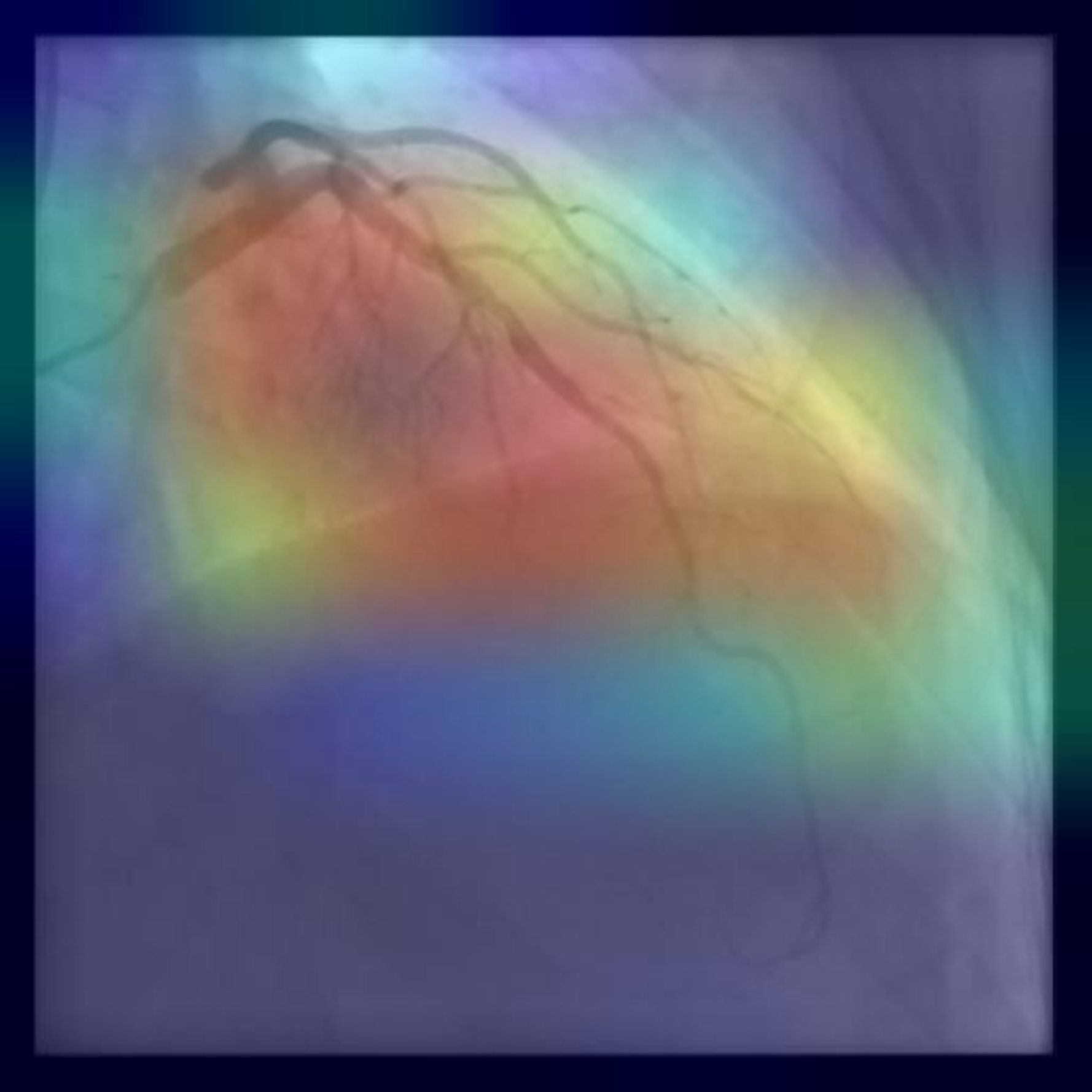}}
    \caption{Grad-CAM visualizations in the RCA and LCA viewing angles showcasing the regions of the frame that most contribute to their correct classification.}
    \label{res:viewsgradcam}
\end{figure}
To better understand which regions the model is focusing on the frame to decide the correct viewing angle, gradient-weighted class activation maps (Grad-CAM) \cite{sota:gradCam} were employed to visualize the degree of contribution of specific image regions. 
It is possible to observe (see \autoref{res:viewsgradcam}) that the larger 512 by 512-pixel resolution model, by having more parameters and larger feature map dimensions does not focus on the coronary artery themselves to differentiate the designated viewing angle. Instead, it focuses on more fine-grained patterns of the human morphology, as a result from the variations of the C-arm X-ray unit. On the other hand, the 224 by 224 resolution model, due to the scaled-down resolution (resulting in lower-dimensional feature maps), captures the more broad patterns of the LCA while still detecting human morphology patterns in the RCA.
\begin{table}[!htp]
\centering
\caption{Coronary viewing angles classification metrics performance.}\label{res:viewsComp}
\begin{tabular}{cccc}
\hline
Image Dimensions & Accuracy & F1 Score & Cross Entropy \\ \hline
512 & 0.96$\pm$0.01 & 0.96$\pm$0.01 & 0.14$\pm$0.27 \\
224 & 0.97$\pm$0.01 & 0.97$\pm$0.01 & 0.08$\pm$0.31 \\ \hline
\end{tabular}
\end{table}
From the observed performance shown in \autoref{res:viewsComp}, it is clear that the model can correctly relate the images to their respective viewing angles. This is important since decreased dimensions significantly improve training and inference time. The scaled-down image input model shows marginal increases in accuracy and F1 score but a considerably lower value in the cross-entropy loss, which corresponds to more confidence in the viewing angles predictions.

\subsection{Stenosis detection performance}

For the stenosis detection task, the objective is to generate bounding box proposals with a high IoU and confidence score with reference to ground truth annotations. Our detection model is configured to output a maximum of 5 bounding boxes at inference time. For recall and precision evaluation, a detection is considered a true positive if the IoU is greater than 0.2 and confidence score above 0.5. Additionally, the performance of at least one candidate bounding box per sequence, corresponding to a ground truth, is also shown.
 \begin{figure*}[!ht]
    \centering
    \includegraphics[width=\linewidth]{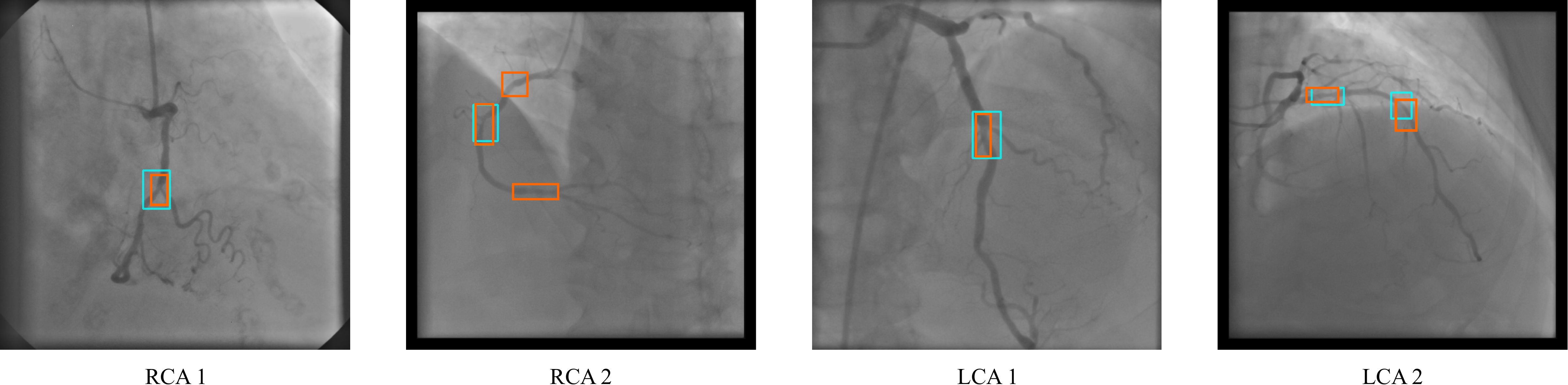}
    \caption{Stenosis detection examples in validation set for RCA and LCA viewing angles with cyan bounding boxes denoting ground truth annotations and orange representing the estimated ones.}
    \label{resstenDetExamples}
\end{figure*}
Our baseline (\textit{B}) is defined as having all frames from the full radio-opaque contrast interval included in training. Performance is evaluated only in reference frames. In attempts to improve the model's capacity at differentiating positive examples (stenosis) from negative ones (background/healthy coronaries), experiments were made including background (\textit{BG}, frames without any contrast) and healthy coronary frames (\textit{NL}) to the RCA and LCA models.
\begin{table}[!ht]
\centering
\caption{Comparison of stenosis detection metrics with previous work}\label{res:detectionComp}
\resizebox{\linewidth}{!}{\begin{tabular}{ccccccc}
\hline
                     &             &              & \multicolumn{2}{c}{max 1 det @ 0.2 IoU} & \multicolumn{2}{c}{max 5 dets @ 0.2 IoU} \\
                     &             & At least One & Recall            & Precision           & Recall              & Precision          \\ \hline
\multirow{5}{*}{RCA} & B           & \textbf{0.81±0.01}    & \textbf{0.72±0.03 }        & 0.80±0.02        & \textbf{0.73±0.01 }       & 0.64±0.04       \\
                     & BG          & 0.74±0.03    & 0.64±0.03         & 0.714±0.05        & 0.61±0.05        & 0.64±0.01       \\
                     & NL          & 0.74±0.04    & 0.68±0.04         & \textbf{0.82±0.02}         & 0.51±0.04        & \textbf{0.72±0.03 }      \\
                     & BGNL        & 0.71±0.02    & 0.65±0.03         & 0.79±0.03        & 0.63±0.02         & 0.71±0.02       \\
                     & Cong et al. & -            & 0.71              & -                   & -                   & -                  \\ \hline
\multirow{5}{*}{LCA} & B           & \textbf{0.77±0.03}    & 0.68±0.04         & 0.79±0.03        & 0.65±0.01         & 0.65±0.01       \\
                     & BG          & 0.74±0.04    & \textbf{0.70±0.04 }        & \textbf{0.84±0.02}        & \textbf{0.68±0.03 }       & \textbf{0.74±0.02}       \\
                     & NL          & 0.71±0.02    & 0.65±0.02         & 0.81±0.03        & 0.56±0.01        & 0.68±0.01       \\
                     & BGNL        & 0.65±0.03    & 0.58±0.03         & 0.73±0.02        & 0.51±0.02        & 0.64±0.01       \\
                     & Cong et al. &              & 0.60              & -                   & -                   & -                  \\ \hline
\end{tabular}}
\end{table}
We compare the performance of all our models against themselves and with previous work. \autoref{res:detectionComp} shows that our model outperforms the model of Cong et al. \cite{AI:sten4}. 
From visual validation of the model's performance (see \autoref{resstenDetExamples}), it is possible to observe that it performs better in frames with only one stenosis. However, in cases where more than one stenosis is present, the model struggles at the detection in its entirety. Nevertheless, the models can detect several stenoses per frame even with hard examples (iFR above threshold), achieving good performances.
The results show that our variations on the base model improve the performance on the LCA, but the addition of more true negative examples on the RCA did not boost performance.
\section{Conclusions and Future Work}\label{sec:conclusions}
The aim of this work was to identify and assess the severity of stenosis from X-ray coronary angiographies.
A two-stage framework based on convolutional neural networks was assembled to automate the detection of stenosis location. The first objective was to precisely classify reference frames as belonging to the right coronary artery (RCA) or to the left coronary artery (LCA). High performance metrics of 0.97 accuracy and 0.97 F1 score were obtained with transfer learning and fine-tuning of the ResNet-50.

After detecting the RCA and LCA viewing angles, two distinct models, based on the single shot detector RetinaNet architecture were assembled as the second stage of the framework to automatically detect stenosis location. Comparisons with different authors confirm the superior performance of our method, which obtains scores of 0.72/0.70 recall for one detection, 0.73/0.68 recall and 0.72/0.74 precision for 5 detection on the RCA/LCA, respectively. Our models performed reasonably well in the detection of single and multiple stenosis but they still leave considerable room for improvement, as many background regions were also detected as stenosis. We will publicly release the dataset used in this work, once the appropriate clearances are obtained.

In what regards future work, it is important to estimate the iFR from the detected regions of interest, an area we are actively developing. We believe that attention mechanisms will be useful in the improvement of the bounding box detection and in the estimation of the iFR values from the detected bounding boxes, since stenosis only represents a small portion of the frame.
\bibliographystyle{splncs04}
%




\bibliography{bibliography.bib}

\begin{thebibliography}{10}
\providecommand{\url}[1]{\texttt{#1}}
\providecommand{\urlprefix}{URL }
\providecommand{\doi}[1]{https://doi.org/#1}

\bibitem{sota:stenDet2018}
Antczak, K., Liberadzki, L.: Stenosis detection with deep convolutional neural
  networks. MATEC Web of Conferences  \textbf{210},  04001 (Jan 2018)

\bibitem{AI:sten1}
Au, B., Shaham, U., Dhruva, S., Bouras, G., Cristea, E., Lansky, A., Coppi, A.,
  Warner, F., Li, S., Krumholz, H.M.: Automated characterization of stenosis in
  invasive coronary angiography images with convolutional neural networks. CoRR
   \textbf{abs/1807.10597} (2018)

\bibitem{sota:mri}
Avendi, M., Kheradvar, A., Jafarkhani, H.: A combined deep-learning and
  deformable-model approach to fully automatic segmentation of the left
  ventricle in cardiac mri. Medical Image Analysis  \textbf{30},  108--119
  (2016)

\bibitem{AI:sten4}
{Cong}, C., {Kato}, Y., {Vasconcellos}, H.D., {Lima}, J., {Venkatesh}, B.:
  Automated stenosis detection and classification in x-ray angiography using
  deep neural network. In: 2019 IEEE International Conference on Bioinformatics
  and Biomedicine (BIBM). pp. 1301--1308 (2019)

\bibitem{imp:fastrcnn}
{Girshick}, R.: Fast r-cnn. In: 2015 IEEE International Conference on Computer
  Vision (ICCV). pp. 1440--1448 (2015)

\bibitem{imp:boxPar}
{Girshick}, R., {Donahue}, J., {Darrell}, T., {Malik}, J.: Rich feature
  hierarchies for accurate object detection and semantic segmentation. In: 2014
  IEEE Conference on Computer Vision and Pattern Recognition. pp. 580--587
  (2014)

\bibitem{imp:resnetpaper}
{He}, K., {Zhang}, X., {Ren}, S., {Sun}, J.: Deep residual learning for image
  recognition. In: 2016 IEEE Conference on Computer Vision and Pattern
  Recognition (CVPR). pp. 770--778 (2016)

\bibitem{imp:adam}
Kingma, D.P., Ba, J.: Adam: {A} method for stochastic optimization. In: Bengio,
  Y., LeCun, Y. (eds.) 3rd International Conference on Learning
  Representations, {ICLR} 2015, San Diego, CA, USA, May 7-9, 2015, Conference
  Track Proceedings (2015)

\bibitem{sota:mri3}
{Li}, Z., {Lin}, A., {Yang}, X., {Wu}, J.: Left ventricle segmentation by
  combining convolution neural network with active contour model and tensor
  voting in short-axis mri. In: 2017 IEEE International Conference on
  Bioinformatics and Biomedicine (BIBM). pp. 736--739 (2017)

\bibitem{imp:fpn}
{Lin}, T., {Dollár}, P., {Girshick}, R., {He}, K., {Hariharan}, B.,
  {Belongie}, S.: Feature pyramid networks for object detection. In: 2017 IEEE
  Conference on Computer Vision and Pattern Recognition (CVPR). pp. 936--944
  (2017)

\bibitem{imp:retinanet}
{Lin}, T., {Goyal}, P., {Girshick}, R., {He}, K., {Dollár}, P.: Focal loss for
  dense object detection. In: 2017 IEEE International Conference on Computer
  Vision (ICCV). pp. 2999--3007 (2017)

\bibitem{meddata:tracker}
Lukežič, A., Vojíř, T., Čehovin Zajc, L., Matas, J., Kristan, M.:
  Discriminative correlation filter tracker with channel and spatial
  reliability. International Journal of Computer Vision  \textbf{126}(7),
  671–688 (Jan 2018)

\bibitem{Bio:CADd}
National Heart, Lung, and Blood Institute: Ischemic Heart Disease (2013),
  https://www.nhlbi.nih.gov/health-topics/ischemic-heart-disease

\bibitem{sota:mri4}
Poudel, R.P.K., Lamata, P., Montana, G.: Recurrent fully convolutional neural
  networks for multi-slice mri cardiac segmentation. In: Zuluaga, M.A., Bhatia,
  K., Kainz, B., Moghari, M.H., Pace, D.F. (eds.) Reconstruction, Segmentation,
  and Analysis of Medical Images. pp. 83--94. Springer International
  Publishing, Cham (2017)

\bibitem{AI:yolo}
Redmon, J., Divvala, S., Girshick, R., Farhadi, A.: You only look once:
  Unified, real-time object detection. In: 2016 IEEE Conference on Computer
  Vision and Pattern Recognition (CVPR). pp. 779--788 (2016)

\bibitem{AI:unet}
Ronneberger, O., Fischer, P., Brox, T.: U-net: Convolutional networks for
  biomedical image segmentation. In: Navab, N., Hornegger, J., Wells, W.M.,
  Frangi, A.F. (eds.) Medical Image Computing and Computer-Assisted
  Intervention -- MICCAI 2015. pp. 234--241. Springer International Publishing,
  Cham (2015)

\bibitem{meddata:osirix}
Rosset, A., Spadola, L., Ratib, O.: Osirix: An open-source software for
  navigating in multidimensional dicom images. Journal of digital imaging : the
  official journal of the Society for Computer Applications in Radiology
  \textbf{17},  205--16 (Oct 2004)

\bibitem{Coro:stat}
Roth, G.A., Abate, D., Abate, K.H., et~al.: Global, regional, and national
  age-sex-specific mortality for 282 causes of death in 195 countries and
  territories, 1980–2017: a systematic analysis for the global burden of
  disease study 2017. The Lancet  \textbf{392}(10159),  1736 -- 1788 (2018)

\bibitem{sota:gradCam}
{Selvaraju}, R.R., {Cogswell}, M., {Das}, A., {Vedantam}, R., {Parikh}, D.,
  {Batra}, D.: Grad-cam: Visual explanations from deep networks via
  gradient-based localization. In: 2017 IEEE International Conference on
  Computer Vision (ICCV). pp. 618--626 (2017)

\bibitem{imp:inceptionv3}
{Szegedy}, C., {Vanhoucke}, V., {Ioffe}, S., {Shlens}, J., {Wojna}, Z.:
  Rethinking the inception architecture for computer vision. In: 2016 IEEE
  Conference on Computer Vision and Pattern Recognition (CVPR). pp. 2818--2826
  (2016)

\bibitem{sota:mri2}
Tan, L.K., Liew, Y.M., Lim, E., McLaughlin, R.A.: Convolutional neural network
  regression for short-axis left ventricle segmentation in cardiac cine mr
  sequences. Medical Image Analysis  \textbf{39},  78 -- 86 (2017)

\bibitem{sota:stenDet2020}
Wu, W., Zhang, J., Xie, H., Zhao, Y., Zhang, S., Gu, L.: Automatic detection of
  coronary artery stenosis by convolutional neural network with temporal
  constraint. Computers in Biology and Medicine  \textbf{118},  103657 (2020)

\bibitem{mri:5}
{Zreik}, M., {Leiner}, T., {de Vos}, B.D., {van Hamersvelt}, R.W., {Viergever},
  M.A., {Išgum}, I.: Automatic segmentation of the left ventricle in cardiac
  ct angiography using convolutional neural networks. In: 2016 IEEE 13th
  International Symposium on Biomedical Imaging (ISBI). pp. 40--43 (2016)

\end{thebibliography}
\end{document}